\documentclass[prb,aps,floatfix, twocolumn]{revtex4}
\usepackage{graphicx,bm}
\usepackage{subfigure}
\usepackage{amssymb}
\usepackage{psfrag}
\usepackage{amsmath}
\usepackage{verbatim}
\usepackage{my_symbols}
\usepackage{amsfonts}
\usepackage{graphicx}

\begin{document}

\title{Spin-transfer torque in magnetic tunnel junctions: Scattering theory}

%

\author{Jiang Xiao, Gerrit E.W. Bauer} \affiliation{Kavli Institute of NanoScience, Delft
University of Technology, 2628 CJ Delft, The Netherlands} 

\author{Arne Brataas} \affiliation{Department of Physics, Norwegian University of Science and
Technology, NO-7491 Trondheim, Norway}

\begin{abstract}

We study the bias-dependent spin-transfer torque in magnetic tunnel junctions in the Stoner
model by scattering theory. We show that the in-plane (Slonczewski type) torque vanishes and
subsequently reverses its direction when the bias voltage becomes larger or the barrier wider
than material and device-dependent critical values. We are able to reproduce the magnitude and
the bias dependence of measured in-plane and out-of-plane torques using realistic parameters.
The condition for the vanishing torque is summarized by a phase diagram depending on the
applied bias and barrier width, which is explained in terms of an interface spin polarization
and the electron focusing by the barrier. Quantum size effects in the spin-transfer torque are
predicted as a function of the thickness of a normal metal layer inserted between the
ferromagnet and tunnel barrier.

\end{abstract} 

\date{\today} \maketitle

\pagestyle{plain}

\section{Introduction} \label{sec:intro}

Magnetic tunnel junctions (MTJ) are layered structures in which an insulating tunnel
barrier (I) separates two ferromagnetic layers (F). \cite{Zhang:2003, Tsymbal:2003} The
interplay between electronic currents and an order parameter difference, i.e.
magnetizations rotated away from the equilibrium configurations, is the magnetic
equivalent to the Josephson effect in superconductivity. The MTJ with a thin normal metal
insertion layer is the only magneto-electronic structure that shows quantum size effects
on electron transport. \cite{Yuasa:2002} MTJs based on epitaxial MgO barriers
\cite{Yuasa:2004, Parkin:2004} are used in the magnetic random-access memory (MRAM)
devices that are operated by the spin-transfer torque. \cite{Slonczewski:1996,
Berger:1996} MTJs have been studied vigorously, initially focusing on the tunnel
magneto-resistance (TMR). \cite{Slonczewski:1989, Miyazaki:1995, Moodera:1995, Yuasa:2002}
More recently, the focus shifted to spin-transfer torque and current-induced magnetization
switching. \cite{Huai:2004, Slonczewski:2005, Kalitsov:2006, Levy:2006, Theodonis:2006,
Petit:2007, Heiliger:2008, Manchon:2007c, Wilczynski:2008, Manchon:2008, Sankey:2008,
Kubota:2008, Deac:2008, Sun:2008} On the theoretical front, spin-transfer effects have
been studied extensively in metallic spin valve structures based on various models.
\cite{Brataas:2000, Waintal:2000, Slonczewski:2002, Stiles:2002b, Xiao:2007,
Slonczewski:2007} For tunneling structures, such studies are still relatively scarce.
\cite{Theodonis:2006, Heiliger:2008, Manchon:2007c, Wilczynski:2008, Manchon:2008} 


Here we report a model study of the spin-transfer torque in magnetic tunnel junctions.  Since
the ferromagnets are separated by tunnel barriers, we cannot use theories existing for metallic
structures that are mostly based on semiclassical methods.  \cite{Brataas:2000,
Tserkovnyak:2005} Instead, we chose a fully quantum mechanical treatment of transport through
the tunnel barrier by scattering theory. The high quality of MgO tunnel junctions and the
prominence of quantum oscillations observed in FNIF structures (even for alumina barriers)
provide the motivation to concentrate on ballistic structures in which the transverse Bloch
vector is conserved during transport. We qualitatively (and even quantitatively) confirm the
results in Ref.  \onlinecite{Theodonis:2006, Wilczynski:2008, Manchon:2008}. However, our model
is simpler and physically more transparent than the tight-binding method used in Ref.
\onlinecite{Theodonis:2006} and the numerical studies of Refs.  \onlinecite{Wilczynski:2008}
and \onlinecite{Manchon:2008}. We are able to reproduce simultaneously both the in-plane and
out-of-plane torque experimental data using realistic material and device parameters, in
contrast to a fit based on the tight-binding model. \cite{Kubota:2008} We also show finite
zero-bias out-of-plane torque for asymmetric structures. Scattering theory enables us to
distill a clear physical picture of the peculiarities of the spin-transfer torque in MTJs,
which allows us to understand why and when the torque goes to zero. The torque zero-crossing
condition can be summarized by a phase diagram spanned by the applied bias and barrier width
parameters. With our approach we can go beyond the FIF MTJ and study the effect of a normal
metal insertion (FINF structures). Quantum size oscillations in the torque are predicted as a
function of the thickness of the N insertion layer. 

This paper is organized as follows: Section \ref{sec:sm} introduces the FIF and FNIF structures
and the scattering theory. In Section \ref{sec:approx}, approximations are introduced in order
to derive analytic expressions. Section \ref{sec:res} presents our main results. Section
\ref{sec:comp} compares our model with experimental results. Section \ref{sec:dis} and
\ref{sec:sum} contain a brief discussion and summary, respectively. Two appendices are attached
at the end.

\section{Structure \& Method} \label{sec:sm}

We consider multilayers as shown in \Figure{fig1}(a) in which two semi-infinite F leads
(F(L) and F(R)) are connected by an insulating layer (I) of width $d$ and a non-magnetic
metal layer (N) of width $a$. The magnetization direction of F(L)/F(R), $\ml/\mr$
($\abs{\ml}=\abs{\mr}=1$), is treated as fixed/free. This structure reduces to a
conventional FIF MTJ when $a=0$. 

Let $A, B, C, D, C', D', E, F$ be the spin-dependent {\it amplitudes} ($A^\dagger =
(A_\up^\dagger, A_\dn^\dagger)$) of flux-normalized spinor wave-functions at specific
points. The scattering states can be expressed in terms of two incoming waves $A$ and $F$,
such as: 
\begin{equation}
	C' = \hat{s}_{\rm C'A} A + \hat{s}_{\rm C'F} F, 
\end{equation}
where $\hat{s}_{\rm C'A}$ and $\hat{s}_{\rm C'F}$ are $2\times 2$ matrices in spin space
that can be constructed by concatenating the scattering matrices of region $S_{1,2}$ and
of the insulating layer bulk (see \Figure{fig1}). To leading order of the transmission
($t_{\rm b}$) through the bulk I (similar expansions hold for $\hat{s}_{\rm D'A}$ and
$\hat{s}_{\rm D'F}$ as well):
\begin{figure}[t]
	\includegraphics[width=8.5cm]{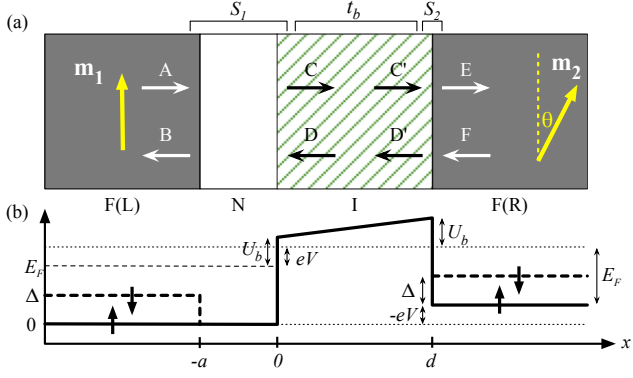}
	 \caption{(Color online) (a): FNIF heterostructure, in which $S_{1,2}$ indicate two
	 different interface scattering regions; (b): The potential profiles (at positive
	 bias) for majority and minority electron spins in F are shown by solid and dashed
	 lines respectively. The exchange splitting is $\Delta$ and the tunnel barrier has
	 height $U_{\rm b}$ relative to the Fermi energy $\EF$. The applied bias $V$ pictured
	 in (b) corresponds to a net electron (particle) flow from right to left.} 
	\label{fig1}
\end{figure}
\begin{subequations}
\label{eqn:sC}
\begin{align}
	\hat{s}_{\rm C'A} &= \midb{(1-r_{\rm b}'\hat{r}_2)^{-1} t_{\rm b}
							   (1-\hat{r}_1'r_{\rm b})^{-1} }\hat{t}_1, \\
	\hat{s}_{\rm C'F} &= \left[ (1-r_{\rm b}'\hat{r}_2)^{-1}r_{\rm b}' \right. \nn
							 &+\left.t_{\rm b} (1-\hat{r}_1'r_{\rm b})^{-1} \hat{r}_1'
							   t_{\rm b}'(1-\hat{r}_2r_{\rm b}')^{-1} \right] \hat{t}_2', 
\end{align}
\end{subequations}
where $\hat{t}_{1,2}$/$\hat{r}_{1,2}$ are the $2\times 2$ transmission/reflection matrices
for $S_{1,2}$ (see \Figure{fig1}) and $t_{\rm b}/r_{\rm b}$ are the spin-independent
transmission/reflection coefficient for the insulating bulk material that are proportional
to the unit matrix in spin space and therefore without hat. The primed and unprimed
versions indicate scattering of electrons impinging from the left and right, respectively.
The reflection coefficient $r_{\rm b}$ is due to the impurity scattering inside the bulk
insulator, which, as $t_{\rm b}$, contains an exponential decay factor representing
evanescent states in I. For this reason the magnitude of $r_{\rm b}$ is comparable to or
much smaller than that of $t_{\rm b}$ depending on the density of the impurities. All
scattering matrices are matrices in ${\bf k}$-space defined by the propagating states of
left and right leads in the energy window available for transport, labeled by their
transverse wave vectors in the leads: ${\bf q, q'}$ (the band index is suppressed) at a
given energy. 

An applied bias voltage $V$ drives a (conserved) charge current $J_{\rm c}$ and
(spatially-dependent) spin current ${\bf J}_{\rm s}$ through the device. At zero
temperature, the charge current reads,
\begin{align}
	J_{\rm c} 
	&= {1\over (2\pi)^3}\int dE~\sum_{\bf q, q'}
	~ j_{\rm c}({\bf q, q'}), \label{eqn:Jc}\\
	j_{\rm c} &= {4e\over\hbar} \Trs{\im{
	 \hat{s}_{\rm EA} \hat{s}_{\rm EA}^\dagger f_{\rm L} 
	-\hat{s}_{\rm EF} \hat{s}_{\rm EF}^\dagger f_{\rm R} }}. \nonumber
\end{align}
where $\Trs{\cdots}$ denotes the spin trace, and the summation is over all the transverse
modes at energy $E$. $f_{\rm L} = f_{\rm L}(E)$ and $f_{\rm R} = f_{\rm R}(E+eV)$ are
(zero temperature) Fermi-Dirac electron distribution functions in the left and right
reservoirs. We are therefore disregarding any spin-accumulation in the ferromagnet, which
is valid for tunnel junctions of current interest in which the spin-flip rate in the
ferromagnet is larger than the tunnel rate. The scattering matrices depend on $V$ by the
bias-induced potential profile.  The spin current, or the angular momentum current, at the
left side of the I/F(R) interface (within I) reads 
\begin{align}
	{\bf J}_{\rm s} 
	&= {1\over (2\pi)^3}\int dE~\sum_{\bf q, q'}
	~ {\bf j}_{\rm s}(E, {\bf q, q'}), \label{eqn:Js}\\
	{\bf j}_{\rm s} &=  2\Trs{\hbmsigma~\im{
	 \hat{s}_{\rm C'A} \hat{s}_{\rm D'A}^\dagger f_{\rm L} 
	-\hat{s}_{\rm C'F} \hat{s}_{\rm D'F}^\dagger f_{\rm R} }}. \nonumber
\end{align}


Since the spin current deep in the FM lead is longitudinal to the magnetization, the
torque ${\bf N}$ acting on F(R) is equal to the transverse component of the incoming spin
current that is absorbed at the interface: \cite{Slonczewski:1996, Brataas:2000,
Waintal:2000, Stiles:2002b}
\begin{equation}
	{\bf N} = {\bf J}_{\rm s} - ({\bf J}_{\rm s}\cdot\mr) \mr
	= {\bf N}_\| + {\bf N}_\perp,
	\label{eqn:NJ}
\end{equation}
with the in-plane (Slonczewski) torque ${\bf N}_\| \propto \mr\times(\ml\times\mr)$ and
out-of-plane (field-like) torque ${\bf N}_\perp \propto \ml\times\mr$. Similarly ${\bf n} =
{\bf j}_{\rm s} - ({\bf j}_{\rm s}\cdot\mr)\mr = {\bf n}_\| + {\bf n}_\perp$.

At low bias, the non-equilibrium part of the spin current is proportional to the bias voltage
${\bf J}_{\rm s} - {\bf J}_{\rm s}^0 = {\bf G}_{\rm s}V$, where ${\bf J}_{\rm s}^0$ is the
equilibrium spin current that is related to interlayer exchange coupling at equilibrium, and
${\bf G}_{\rm s}$ is the spin conductance: 
\begin{align}
	{\bf G}_{\rm s}
	&={1\over (2\pi)^2}\sum_{\bf q,q'}^{E=\EF}
	~ {\bf g}_{\rm s}({\bf k, k'}), 
	\label{eqn:Gs}\\
	{\bf g}_{\rm s} &=  {e\over \pi}~\Trs{\hbmsigma~\im{
	 \hat{s}_{\rm C'A} \hat{s}_{\rm D'A}^\dagger
	+\hat{s}_{\rm C'F} \hat{s}_{\rm D'F}^\dagger }}, \nonumber
\end{align}
where the scattering matrices are evaluated at zero bias $V = 0$, and the summation is over
transport channels at the Fermi energy. We define the linear-response torkance
\cite{Slonczewski:2007} ${\bf T} = {\bf N}/V$, and ${\bf T} = {\bf G}_{\rm s} - ({\bf G}_{\rm
s}\cdot\mr)\mr = {\bf T}_\| + {\bf T}_\perp$ and $\btau = {\bf g}_{\rm s} - ({\bf g}_{\rm
s}\cdot\mr)\mr = \btau_\| + \btau_\perp$.

\section{Approximations} \label{sec:approx}

We assume in the following that the spin is conserved during the scattering. Then $\hat{t}_i$
for $S_i$ ($i = 1, 2$, similar for $\hat{r}_i$) is diagonal when choosing $\mi$ as spin
quantization axis: Expanded in Pauli matrices $\hbmsigma = (\hsigma_x, \hsigma_y, \hsigma_z)$,
$\hat{t}_i = t_i^+ + t_i^-\hbmsigma\cdot\mi$, with $t_i^\pm = (t_i^\up \pm t_i^\dn)/2$.
$t_{i}^\sigma$ ($\sigma = \up, \dn$) is the transmission amplitude for spin-$\sigma$ for spin
quantization axes $\mi$ through the scattering region $S_i$. In the absence of impurities
($r_{\rm b} = 0$) and to leading order of $t_{\rm b}$:
\begin{subequations}
\begin{align}
	\hat{s}_{\rm C'A} &=  t_{\rm b}(t_1^+ + t_1^-\hbmsigma\cdot\ml), \\
	\hat{s}_{\rm D'A} &= (r_2^+t_{\rm b} t_1^+ + r_2^-t_{\rm b} t_1^-\ml\cdot\mr) \\
	&+ \hbmsigma\cdot(r_2^-t_{\rm b} t_1^+\ml + r_2^+t_{\rm b} t_1^-\mr -ir_2^-t_{\rm b} t_1^-\ml\times\mr),\nonumber
\end{align}
\label{eqn:sCA}
\end{subequations}
Similar expansions hold for $\hat{s}_{\rm C'F}$ and $\hat{s}_{\rm D'F}$. 

Next, we adopt the free electron approximation tailored for transition metal ferromagnets.
\cite{Slonczewski:1989} We assume spherical Fermi surfaces for spin-up and spin-down electrons
(in both F(L) and F(R)) with Fermi wave-vectors $\kF^{\up} = \sqrt{2m\EF/\hbar^{2}}$ and
$\kF^{\dn} = \sqrt{2m(\EF-\Delta)/\hbar^{2}}$, with an effective electron mass $m$ in F.
Electrons in N are assumed to be ideally matched with the majority electrons in F
($\kF=\kF^{\up}$). The effective electron mass in the tunnel barrier is assumed to be $m_{\rm
b} = \beta m$. \Figure{fig1}(b) shows the adopted potential profile with barrier height $U_{\rm
b}$. We assume an applied potential bias that is smaller than the barrier height $\abs{eV} <
U_{\rm b}$ that fully drops over the tunnel barrier. Positive bias corresponds to charge
current flow from left to right and electron particle flow from right to left. As in Refs.
\onlinecite{Theodonis:2006, Manchon:2007c, Wilczynski:2008}, we assume that energy and
transverse wave-vector ${\bf q}$ are conserved, thus disregard any impurity/interface roughness
scattering. All scattering matrices then become diagonal in ${\bf k}$-space. 


In the free electron model, the flux-normalized wave-functions in N and F are: 
\begin{equation}
	\psi_\pm^{\rm N} = {e^{\pm ik_x x}\over \sqrt{k_x}} \quad \mbox{in N} \qand
	\psi_\pm^\sigma = {e^{\pm ik_x^\sigma x}\over \sqrt{k_x^\sigma}}
	\quad \mbox{in F},
	\label{eqn:psi}
\end{equation}
Using the WKB approximation, \cite{Manchon:2007c} the wave-function in the tunneling barrier is
\begin{equation}
	\psi_\pm^{\rm b} = {e^{\pm\int_0^x\kappa(w)dw}\over \sqrt{\pm i\kappa(x)}}
	\quad\mbox{in I}
\end{equation}
with
\begin{equation}
	\kappa(x) = \sqrt{{2m_{\rm b}\over\hbar^2}
	\smlb{U_{\rm b} + \EF - E - eV{x\over d}}+q^2},
\end{equation}
where $E$ is the energy of the electron. The WKB approximation is valid when the potential
profile varies slowly in space within the tunneling barrier, {\it i.e.}
$\kappa'(x)\ll\kappa^2(x)$.  The transmission coefficient through I reads $t_{\rm b} =
\exp[-\int_0^d\kappa(w)dw]$.

For finite bias, from \Eq{eqn:Js} and \Eq{eqn:sCA},
\begin{subequations} 
\label{eqn:n}
\begin{align} 
	{\bf n}_\| &= t_{\rm b}^2~
	T_1^-T_2^+(f_{\rm L}-f_{\rm R})\mr\times(\ml\times\mr)
	\label{eqn:nparr}, \\
	{\bf n}_\perp &= 2t_{\rm b}^2~
	\re{T_1^-r_2^-f_{\rm L}+T_2^-{r_1'}^-f_{\rm R}}\ml\times\mr,  
	\label{eqn:nperp} \\
	{\bf n}_\perp^0 &= 2t_{\rm b}^2~
	\re{T_1^-r_2^-+T_2^-{r_1'}^-}f_0\ml\times\mr,  
	\label{eqn:nperp0}
\end{align}
\end{subequations} 
where $T_i^+=\abs{t_i^\up}^2+\abs{t_i^\dn}^2$ is the average transmission probability for
scattering region $S_i$, $T_i^-=p_i T_i^+ = \abs{t_i^\up}^2-\abs{t_i^\dn}^2$ with polarization
$p_i = T_i^-/T_i^+$, and $f_0$ is the equilibrium distribution function at zero bias. ${\bf
n}_\perp$ in \Eq{eqn:nperp} includes both the equilibrium and non-equilibrium contribution to
the out-of-plane torque. The former (${\bf n}_\perp^0$ in \Eq{eqn:nperp0}) is related with the
non-local interlayer exchange coupling. \cite{Slonczewski:1993} The non-equilibrium
contribution is therefore ${\bf n}'_\perp = {\bf n}_\perp - {\bf n}_\perp^0$.  The optical
theorem $2 \im{r_{1,2}^\pm} = T_{1,2}^\pm$ (see Appendix A) is used in the derivation of
\Eq{eqn:nparr} to get rid of all internal reflection in I. In Eq. (11a), we observe that the
in-plane torque is caused by the polarization of the current at the left interface that is
expressed by $T^-_1$. The subsequent absorption of the spin current by the second magnet is
governed by the geometrical projection expressed by the vector product and the total
transparency of the second interface $T^+_2$. It follows from Eq. (11b) that the out
out-of-plane torque has a very different origin. It does not depend directly on the difference
of the electron distributions on both sides of the junctions, but consists of two independent
contributions from both reservoirs. Each contribution consists of the spin-polarization of the
first interface, but is sensitive to the phase of the reflection coefficient of the second
interface. The out-of-plane torque can be interpreted as the net spin created at one interface
that while reflected at the second interface briefly precesses in the exchange field of the
second ferromagnet.

With vanishing bias, $t_{\rm b} = \exp(-\kappa d)$ with $\kappa = \sqrt{2m_{\rm b}U_{\rm
b}/\hbar^2+q^2}$. By \Eq{eqn:Gs} and \Eq{eqn:sCA},
\begin{equation}
	\btau = \btau_\| = {e\over 2\pi}e^{-2\kappa d} T_1^-T_2^+\mr\times(\ml\times\mr),
	\label{eqn:btau}
\end{equation}
and $\btau_\perp = 0$. For reference, the conductance within the same theoretical
framework is given by: \cite{Xiao:CP}
\begin{equation}
	g_{\rm c} = {e^2\over 2h}e^{-2\kappa d}
	\smlb{ T_1^+ T_2^+ + T_1^- T_2^- \ml\cdot\mr}. 
	\label{eqn:gc}
\end{equation}

The vector product $\abs{\mr\times(\ml\times\mr)} = \sin\theta$ in \Eq{eqn:btau} and
$\ml\cdot\mr = \cos\theta$ in \Eq{eqn:gc}, leading to the well-known geometrical dependence of
the angular transport properties of tunnel junctions. \cite{Slonczewski:1996} The vanishing of
the out-of-plane torque, $\btau_\perp = 0$, is a rather general result that holds for symmetric
tunneling junctions and spin valves in the linear-response regime. \cite{Theodonis:2006} We
consider a {\it symmetric} system with an applied voltage $-V/2$ to the left and a voltage
$V/2$ to the right reservoir. To the second order in the bias voltage, the spin-current in the
spacer between the ferromagnets can be expanded as 
\begin{align}
	{\bf I}_s &= 
	\midb{A_1 \ml + B_1 \mr + C_1 \ml \times \mr}V \nn 
	&+ \midb{A_2 \ml + B_2 \mr + C_2 \ml \times \mr}V^2,
	\label{eqn:Is}
\end{align}
When applying the mirror operation left $\leftrightarrow$ right ($1\leftrightarrow2, -V/2
\leftrightarrow V/2, {\bf I}_s \leftrightarrow -{\bf I}_s$) symmetry requires that 
\begin{align}
	-{\bf I}_s &= 
	\midb{A_1 \mr + B_1 \ml + C_1 \mr \times \ml}(-V) \nn 
	&+ \midb{A_2 \mr + B_2 \ml + C_2 \mr \times \ml}(-V)^2,
	\label{eqn:mIs}
\end{align}
which should be identical to \Eq{eqn:Is}. Therefore $A_1=B_1, C_1=0, A_2=-B_2$, whereas
$C_2$ is not restricted. Then the torque on $\mr$ is
\begin{align}
	\NN &= \mr \times ({\bf I}_s \times \mr) \\
	&= \midb{A_1 V -A_2V^2} \mr \times (\ml \times \mr) 
	+ C_2V^2 \ml \times \mr. \nonumber
\end{align}
This proves that, for symmetric systems, the out-of-plane torkance ($\propto 2C_2V$) vanishes
at $V = 0$. It also shows that beyond linear response, there are quadratic (in bias)
contributions to both the in-plane and out-of-plane torque. The argument does not hold for
asymmetric tunneling systems. An experimental zero-bias out-of-plane torkance should therefore
provide interesting information on MTJ asymmetries.


\section{Results} \label{sec:res}

In this section, we discuss three different structures: (A) a symmetric FIF magnetic tunneling
junction, (B) an asymmetric FIF structure in which the left and right FM layers have different
exchange splitting, and (C) an FNIF structure in which a non-magnetic layer is inserted between
the insulator barrier layer and one of the ferromagnetic layers.

\subsection{Symmetric FIF}

For a symmetric Fe/MgO/Fe MTJ: $\kF^{\up} = 1.09$ \r{A}$^{-1}$ and $\kF^{\dn} = 0.42$
\r{A}$^{-1}$ for Fe, \cite{Slonczewski:1989} and $U_{\rm b} \simeq1.0-1.2$ eV and $\beta =
m_{\rm b}/m = 0.4 m$ for MgO.  \cite{Moodera:1996, Bowen:2001, Yuasa:2004, Vincent:2002} This
implies $\EF\simeq4.5$ eV, $\Delta\simeq3.8$ eV $\approx0.85\EF$, and $U_{\rm b}
\approx0.25\EF$.  For an FIF structure ($a = 0$), both $S_1$ and $S_2$ contain only a single
interface.  Using the potential profile in \Figure{fig1}(b), we have
\begin{subequations}
\begin{align}
	t_1^\sigma &= {2\sqrt{i k_1^\sigma\kappa(0)/\beta}\over
	k_1^\sigma+i\kappa(0)/\beta}, &
	t_2^\sigma &= {2\sqrt{i k_2^\sigma\kappa(d)/\beta}\over
	k_2^\sigma+i\kappa(d)/\beta}, \\
	{r_1'}^\sigma &= {-k_1^\sigma+i\kappa(0)/\beta\over
	k_1^\sigma+i\kappa(0)/\beta}, &
	r_2^\sigma &= {-k_2^\sigma+i\kappa(d)/\beta\over
	k_2^\sigma+i\kappa(d)/\beta},
	\label{eqn:t1t2}
\end{align}
\end{subequations}
where $\sigma=\up,\dn$ and $k_1^{\up2} + q^2 = 2mE/\hbar^2$, $k_1^{\dn2} + q^2 =
2m(E-\Delta)/\hbar^2$. $k_2^\sigma$ are defined similarly with $E$ replaced by $E+eV$.  We
set $t_{1,2}^\sigma = 0$ when $\im{k_{1,2}^\sigma}\neq 0$. 


%

\begin{figure}[t]
	\includegraphics[width=8.5cm]{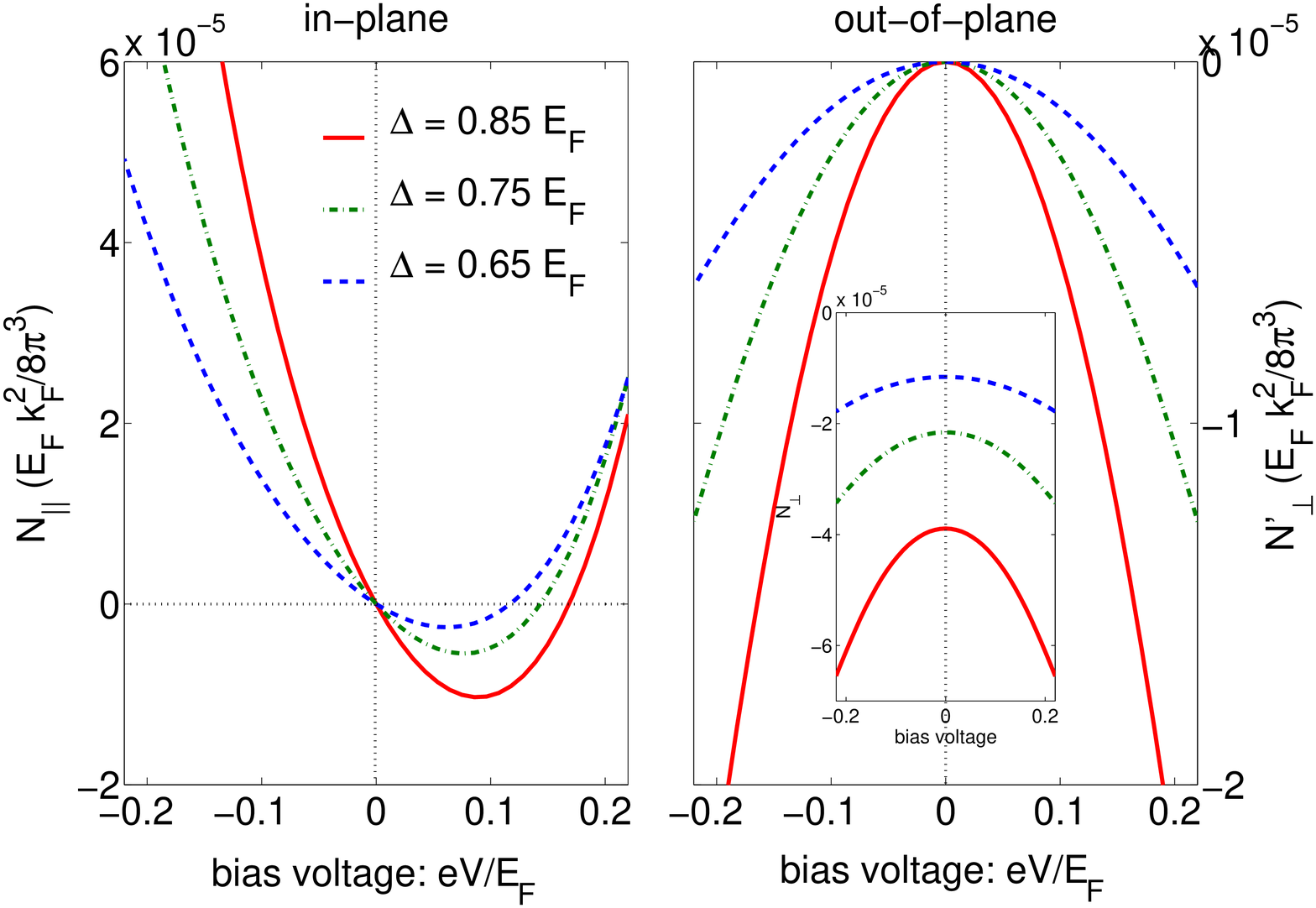}	
	\caption{(Color online) The magnitude of $\bf N_\|$ (left) and $\bf N_\perp'$ (right)
	acting on the right magnetization in \Figure{fig1} vs. applied bias $eV$ for $\beta =
	0.4$ and $d = 1.0$ nm. Inset figure shows the out-of-plane torque including the equilibrium
	contribution at zero bias.}
	\label{N1N2}
\end{figure}

\Figure{N1N2} shows the computed bias dependence of the in-plane torque $\bf N_\|$ (left) and
the non-equilibrium part ({\it i.e.} not containing the equilibrium interlayer exchange
coupling) of the out-of-plane field-like torque ${\bf N}_\perp'(V) = {\bf N}_\perp(V) - {\bf
N}_\perp(0)$ (right) at various exchange splittings for $m_{\rm b} = 0.4 m$ at $d = 1.0$ nm.
The equilibrium exchange coupling gives rise to an effective magnetic field in the LLG
equations that we do not explicitly discuss.  The main features of these curves are: 1) the
in-plane torque has both linear and parabolic contributions; 2) the field-like torque is
parabolic like. These plots are very similar to the corresponding plots by Theodonis {\it et
al}, \cite{Theodonis:2006} meaning that the band structure effects caused by the tight-binding
approximation are not important. The in-plane torque at negative bias and small positive bias
is ``normal'', but changes sign at higher positive bias, where normal means that the direction
of the torque in FIF is the same as the torque in metallic spin valves predicted,
\cite{Slonczewski:1996} {\it i.e.} the torque curve appears in the second and fourth quadrant
in the left panel of \Figure{N1N2}.  When the torque curve is found in the first or third
quadrant, we say that the torque is reversed. In the normal region, the positive bias (electron
flow from right to left) favors the anti-parallel configuration and a negative one the parallel
configuration. In the reversed region, on the other hand, a current polarity that stabilizes
the parallel configuration in the normal region has the opposite effect.

The zero-crossing of the in-plane torque in \Figure{N1N2} can be traced to the sign change of
$T_1^-$ in \Eq{eqn:nparr}, {\it i.e.} the sign change of the polarization of $S_1$ (the F(L)/I
interface)$p_1 = T_1^-/T_1^+ < 0$. \cite{Li-F:2004} The polarization $p_1 \propto \kappa^2(0) -
k_\up k_\dn$, which can take any sign depending on parameters chosen \cite{Slonczewski:2005}
(see Appendix B for a more detailed discussion of this point).  The vanishing torque phenomenon
becomes more transparent without an effective mass mismatch, i.e. for $\beta = 1$ instead of
$\beta = 0.4$ used in \Figure{N1N2}. The polarization vanishes when 
\begin{equation}
	\kappa^2(0) - k_\up k_\dn = 0.
	\label{eqn:k0}
\end{equation}
Since $\kappa(x=0)$ (near left interface) increases and $k_\sigma$ decreases with $q$,
\Eq{eqn:k0} is fulfilled at a certain critical value $q_{\rm c}$. The latter increases
with the electron energy $E$, because $\kappa(0)$ decreases and $k_\sigma$ increases with
$E$. This can be seen clearly from the following equation:
\begin{equation}
	q_{\rm c}^2 = {2m\over\hbar^2}
	\midb{E - {(\EF+U_{\rm b})^2\over 2(\EF+U_{\rm b})-\Delta} }.
	\label{eqn:qc}
\end{equation}
which implies that at the Fermi energy $q_{\rm c}$ is well defined ($q_{\rm c}^2 > 0$)
only when $U_{\rm b}^2<\EF(\EF-\Delta)$.

\begin{figure}[t]
	\psfrag{U}{$U_{\rm b}$}
	\psfrag{E}{$\EF$}
	\psfrag{E2}{$\EF+\abs{eV}$}
	\psfrag{eV}{$\abs{eV}$}
	\psfrag{T1}{$T_1^-$}
	\psfrag{T2}{$T_2^+$}
	\psfrag{negative bias}{negative bias: $V < 0$}
	\psfrag{positive bias}{positive bias: $V > 0$}
	\includegraphics[width=8.5cm]{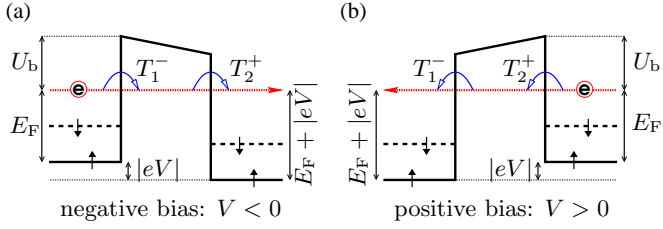}	
	\caption{(Color online) Positive and negative bias situation.}
	\label{bias}
\end{figure}

When $q_{\rm c}^2 < 0$ at low bias there is no polarization sign change for any $q$, and
the (in-plane) torque behaves normally. When the potential profile becomes distorted by an
applied bias as in \Figure{bias}(b), the electrons injected from the right lead have a
maximum energy $E = \EF+eV$. When the applied bias $V$ is large enough, we reach the
regime $q_{\rm c}^2 > 0$, and a polarization sign change of the left interface comes into
play. As $q_{\rm c}$ increases further, more and more electron contribute to the opposite
torque. When $V$ is large enough, the total torque changes sign as seen in \Figure{N1N2}.
On the other hand, when the applied bias is negative (see \Figure{bias}(a)), the transport
is dominated by the electron injected from the left lead. The electron energy and
effective barrier height at the left interface do not change with applied bias, which
means $T_1^-$ (and so the polarization) does not change either. Therefore, we do not see a
zero-crossing (on the right magnetization) at negative bias in \Figure{N1N2}. 

\begin{figure}[b]
	\includegraphics[width=8.5cm]{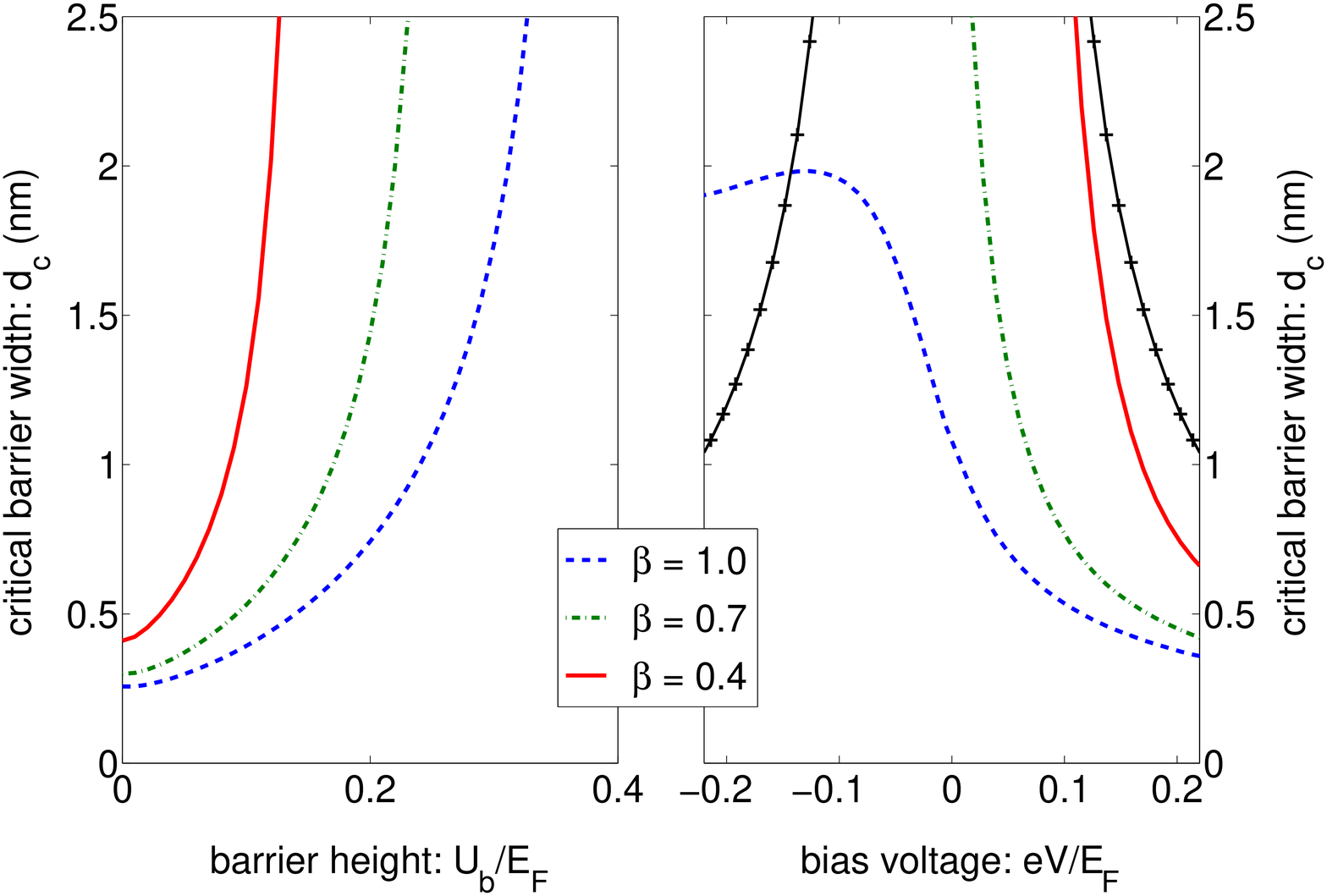}
	\caption{(Color online) Critical barrier width $d_{\rm c}$ vs. barrier height $U_{\rm
	b}$ at linear-response (left) and applied bias $V$ (right). $\Delta = 0.85 \EF$.
	Critical width for TMR at $\beta = 0.4$ (thin solid black line with ``+'' symbols) is
	also shown in the right panel.}
	\label{dc_V}
\end{figure}

We can analogously understand the dependence of the in-plane torque on the barrier width
$d$. We know that there is a polarization sign change for $q < q_{\rm c}$ if $U_{\rm b}$
is not too high (such that $q_{\rm c}^2 > 0$). In tunneling junction, the transport is
dominated by electrons with small $q$ because of the focusing effect due to the
exponential extinction factor in $t_{\rm b}$. When the electron with $q$ values smaller
than $q_{\rm c}$ dominate, the torque reverses sign. At a critical barrier width $d_{\rm
c}$, the contributions from $q<q_{\rm c}$ cancel those from $q>q_{\rm c}$, and the torque
or torkance vanishes: ${\bf N}_\|(d_{\rm c}) = {\bf T}_\|(d_{\rm c}) = 0$ whereas it has
opposite direction for $d > d_{\rm c}$. The left panel of \Figure{dc_V} shows $d_{\rm c}$
vs.  $U_{\rm b}$ in linear-response for various values of the barrier effective mass.
$d_{\rm c}$ increases with $U_{\rm b}$ simply because the polarization sign change
behaviour is less prominent at higher barrier heights ($q_{\rm c}$ decreases with $U_{\rm
b}$). 

The right panel of \Figure{dc_V} shows the bias dependence of $d_{\rm c}$ at barrier height
$U_{\rm b} = 0.25 \EF$ for various barrier effective masses. Because both the positive and
negative bias reduces the average barrier height, $t_{\rm b}$ increases with $\abs{V}$.
However, the applied bias is very ineffective in changing the focusing behaviour, {\it i.e.}
the transmission is hardly less focused by the reduced average barrier height. This is very
unlike the geometrical barrier width $d$, to which the focusing very sensitively: larger
(smaller) $d$ means more (less) focusing. Rather, a positive bias enhances the polarization
sign change behaviour, which leads to a smaller critical barrier width $d_{\rm c}$. Since there
is no polarization sign change ($q_{\rm c}^2 < 0$) at zero or negative bias for $\beta = 0.7$
and $0.4$, a torque zero-crossing is not observed.  For $\beta = 1.0$, polarization could
change sign at zero bias, hence the torque zero-crossing is also observed at negative bias. As
mentioned before, negative bias does not change $T_1^-$ or polarization of the left interface,
but it does change $T_2^+$.  At negative bias, the barrier height at the right interface is
reduced by $\abs{eV}$, which leads to the decrease of $T_2^+$ at small $q$, thus the
polarization sign change behaviour is weakened because of the smaller product $T_1^-T_2^+$ at
small $q$ where the product is negative. Weaker polarization sign change then requires larger
$d_{\rm c}$ for torque zero-crossing at negative bias and we see $d_{\rm c}$ increasing with
negative bias for $\beta = 1.0$. For comparison, a critical barrier width for the sign change
of TMR is also calculated for $\beta = 0.4$, and is shown as the solid black curve with ``+''
symbol in the right panel of \Figure{dc_V}. Since the TMR is symmetric in the applied bias for
the symmetric structures, the curve is also symmetric.

\subsection{Asymmetric FIF}

\begin{figure}[t]
	\includegraphics[height=5.5cm]{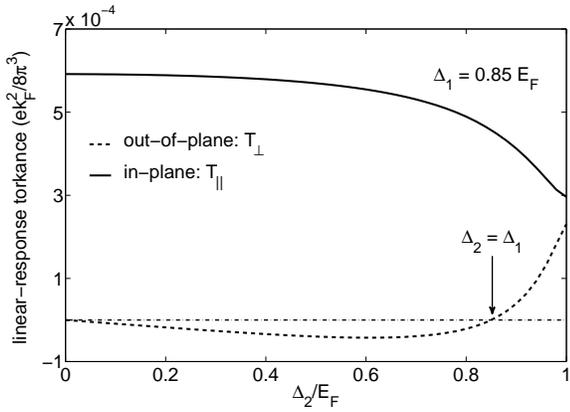}
	\caption{Linear-response torkance vs. exchange splitting $\Delta_2$ in the right FM
	($\Delta_1 = 0.85 \EF, U_{\rm b} = 0.25 \EF, d = 1.0$ nm).}
	\label{ipoop}
\end{figure} 

As discussed at the end of Sec. \ref{sec:approx}, the zero-bias (linear-response) out-of-plane
torkance does not vanish for asymmetric structures. \Figure{ipoop} shows both the in-plane and
out-of-plane torkance for the right FM layer at zero bias ($V = 0$) for an asymmetric FIF
structure, in which the left and the right FM layers have different exchange splitting:
$\Delta_1 = 0.85\EF$ for the left and $\Delta_2$ for the right, where the latter varies from
$0$ to $\EF$. From \Figure{ipoop}, we can see that the out-of-plane torkance is generally
non-zero for asymmetric structures when $\Delta_2 \neq \Delta_1$, and it vanishes when the
right layer becomes non-magnetic ($\Delta_2 = 0$) or when the structure becomes symmetric
($\Delta_2 = \Delta_1$). The in-plane torkance in \Figure{ipoop} decreases with $\Delta_2$
simply because the average transmission probability through the right interface $T_2^+$
decreases. 

\subsection{FNIF}

\begin{figure}[b]
	\includegraphics[height=5.5cm]{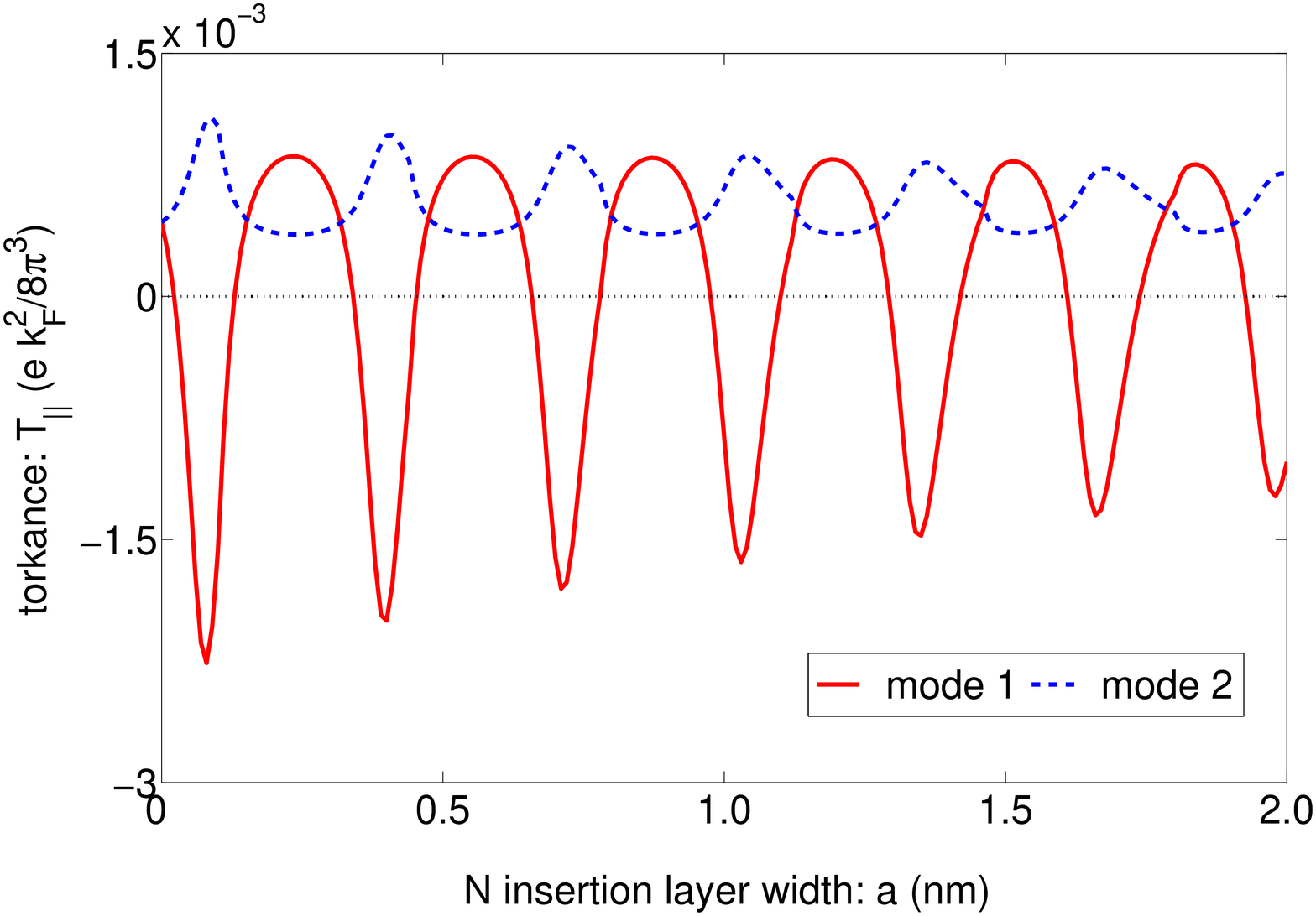}
	\caption{(Color online) Linear-response in-plane torkance vs. non-magnetic insertion layer
	width $a$ ($\Delta = 0.85 \EF, U_{\rm b} = 0.25 \EF, d = 1.0$ nm).}
	\label{torkance-a}
\end{figure} 

An FNIF structure, with a non-magnetic layer of width $a$ between one of the F layers and the
insulating I layer, has never been studied in the regime of spin-transfer torque. Such an
asymmetric FNIF device can be operated in two non-equivalent modes (as in any non-symmetric
MTJ): the left F is static and the right F ($\tilde{\rm F}$) is free (mode 1: FNI$\tilde{\rm
F}$), and vice versa (mode 2: $\tilde{\rm F}$NIF).  Eqs.~(\ref{eqn:n}) - (\ref{eqn:btau}) apply
to mode 1, and apply to mode 2 with subscript 1 and 2 exchanged.  The $a$-dependence of the
in-plane torkance (in linear-response) is shown in \Figure{torkance-a}. The sign of the
in-plane torkance can be controlled by $a$ in mode 1, but not in mode 2. This sign is
determined by the sign of $T_1^-$. In mode 1, $T_1^-(a)$ covers region F(L)-N-I and its sign
can be modulated by the N insertion layer width $a$. However, in mode 2, $T_1^-$ covers I-F(R),
which is independent of $a$, therefore the sign is unchanged. The $a$-dependence of the
in-plane torkance in mode 2 comes from $T_2^+(a)$, which is always positive. Due to the
aliasing effect caused by discrete thickness of the N layer, \cite{Chappert:1991} the period of
the quantum oscillation in \Figure{torkance-a} should be about $\pi/\abs{\kF - \pi/\lambda}$
instead of $\pi/\kF\approx3$ \r{A} shown in the figure, where $\lambda$ is the monolayer
thickness for N layer.

The asymmetry in FNIF structure shall also give rise to a finite linear-response out-of-plane
torkance, which shows similar oscillations as the in-plane torkance in \Figure{torkance-a}. The
magnitude of the zero-bias out-of-plane torkance could be comparable to the in-plane
counterpart.

\section{Comparison with experiments} \label{sec:comp}

\begin{figure}[b]
	\includegraphics[width=8.5cm]{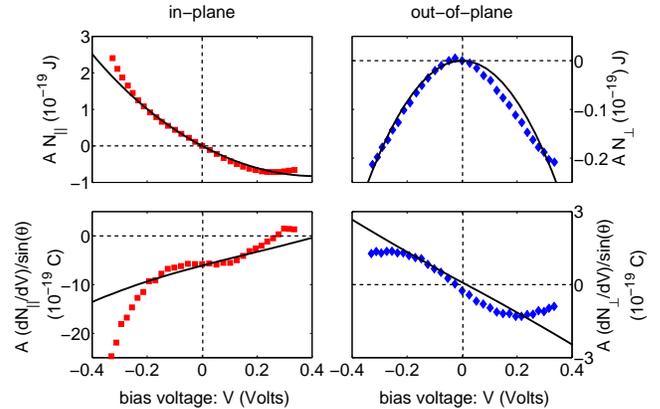}	
	\caption{(Color online) Fitting (solid curve) of the experimental data (dots) from Ref.
	\onlinecite{Kubota:2008}. Top: in-plane torque $A N_\|$ (left) and out-of-plane torque $A
	N_\perp$ (right); Bottom: in-plane torkance $A (dN_\|/dV)/\sin\theta$ (left) and
	out-of-plane torkance $A (dN'_\perp/dV)/\sin\theta$ (right). The following parameters are
	used in all fittings: $\EF = 4.5$ eV, $\Delta = 0.87\EF \approx 3.9$ eV, $U_{\rm b} =
	0.23\EF \approx 1.0$ eV, $\beta = 0.36$; $d = 1$ nm, cross section area $A = 70$ nm$\times
	250$ nm, and $\theta = 137^\circ$ from Ref. \onlinecite{Kubota:2008}.}
	\label{kubota}
\end{figure}

Using realistic material parameters and the geometry parameters provided by Ref.
\onlinecite{Kubota:2008}, we are able to reproduce, even the absolute scale, the experimental
data from Ref. \onlinecite{Kubota:2008} as shown in \Figure{kubota}, which includes the bias
dependence of the in-plane torque ${\bf N}_\|(V)$ and the non-equilibrium part of the
out-of-plane torque ${\bf N}_\perp'(V) = {\bf N}_\perp(V) - {\bf N}_\perp(0)$ and the
corresponding torkance. The experimental data for the torkance in the bottom panels of
\Figure{kubota} are adapted from Fig.~S3(d) ($\beta'_{\rm ST, FT} $) and Fig.~2 (I-V data) of
Ref. \onlinecite{Kubota:2008} by $(dN_{\|,\perp}/dV)/\sin\theta =
(dI/dV)(dN_{\|,\perp}/dI)/\sin\theta = (dI/dV)\beta'_{\rm ST, FT}$. Our model appears to have a
problem with the upturn of the torque at higher positive bias.  The fit in Ref.
\onlinecite{Kubota:2008} based on the tight-binding model of Ref.  \onlinecite{Theodonis:2006}
is slightly better in this respect for a rather large exchange splitting. However, the
out-plane torque is poorly reproduced for the same parameter set.  In contrast, we succeed with
a single set of (realistic) parameters to reproduce both in-plane and out-of-plane torques
(torkance). The resistance for this particular device in our model is $R(\theta = 137^\circ)
\approx 150~\Omega$, which is consistent with the experimental values ($\sim 200~\Omega$).
However, the TMR value in our calculation (about 15\%) is considerably smaller than that in the
experiment.  We believe that the spin-transfer torque is better represented by the free
electron model than the TMR because TMR $\propto p_1p_2$ whereas torque on $\mr \propto p_1$.
If the polarization $p_1, p_2$ at the interfaces are underestimated by a factor of $\eta$, the
TMR value is too small by a factor of $\eta^2$, whereas the spin-transfer torque is affected
only by a factor of $\eta$ (see also Section \ref{sec:dis}). In addition to this, the TMR value
depends sensitively on the exchange splitting $\Delta$. For instance, the TMR value increases
from 15\% to 30\% when $\Delta$ increases from $0.87\EF$ to $0.9\EF$, 

\begin{figure}[b]
	\includegraphics[width=8.5cm]{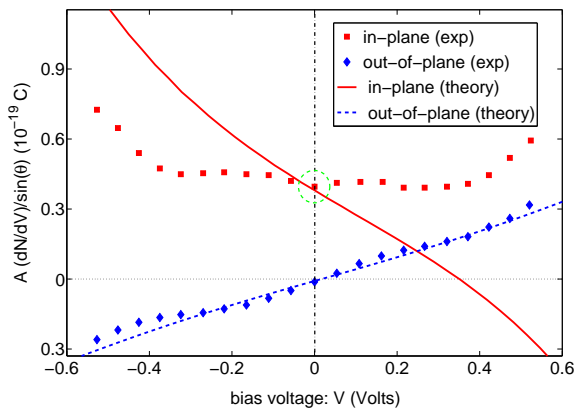}	
	\caption{(Color online) Fitting (curves) of the experimental data (dots) from Ref.
	\onlinecite{Sankey:2008}. The following parameters are used in both fittings: $\EF =
	4.5$ eV, $\Delta = 0.85\EF \approx 3.8$ eV, $U_{\rm b} = 0.25\EF \approx 1.1$ eV,
	$\beta = 0.43$; $d = 1.25$ nm, cross section area $A = 50$ nm$\times 100$ nm, and
	$\theta = 71^\circ$ from Ref. \onlinecite{Sankey:2008} (Notice the sign convention in
	Ref. \onlinecite{Sankey:2008} is opposite to that in Ref. \onlinecite{Kubota:2008}).}
	\label{sankey}
\end{figure}

Another set of the experimental data is shown in \Figure{sankey} adopted from Fig. 3(a) of
Ref. \onlinecite{Sankey:2008}. The experimental data are now the in-plane (red squares)
and out-of-plane (blue diamonds) torkance. We fit the out-of-plane and the linear part of
the in-plane torkance again on absolute scale. Theory contains a quadratic component of
the in-plane torque, which does not show up in this experiment. The resistance for this
particular device in our model is $R(\theta = 71^\circ) \approx 4$ k$\Omega$ consistent
with the experimental value ($\sim 3.5$ k$\Omega$),  whereas our TMR 5\% again is too small.
Note that the torkances are much smaller than in Ref. \onlinecite{Kubota:2008} because of
the thicker barrier and the smaller cross section area, as reflected by the higher
resistance.

Deac {\it et al.} \cite{Deac:2008} also measured the in-plane and out-of-plane torque in a
MgO based tunneling junction. The out-of-plane torque in this experiment agrees well with
other experiments and theory. However, the in-plane torque depends parabolically on the
applied bias ($AV^2$ for positive bias and $-BV^2$ for negative bias, where $A$ and $B$
are positive constants), which is quite different from both Ref. \onlinecite{Sankey:2008}
and \onlinecite{Kubota:2008}. A voltage noise measurements done by Petit {\it et al.}
\cite{Petit:2007} also suggest a linear out-of-plane torkance (or parabolic out-of-plane
torque), which is about 20\% of the in-plane counterpart. Hence all experiments and
theories appear to agree on the out-of-plane torque, whereas consensus about the in-plane
torque has not been reached yet.

\section{Discussion} \label{sec:dis}

Because of the high quality of epitaxial MgO tunnel layers,\cite{Yuasa:2004, Parkin:2004}
we ignored interface roughness and barrier disorder. The main effect of the geometric
interface roughness is to reduce the nominal thickness of the barrier \cite{Zhang:1999}.
Impurity states in the barrier generally increases tunneling because of the opening of
additional tunneling channels with lower barrier height $U_{\rm b}^\prime < U_{\rm b}$.
Impurities states also weaken the spin-dependent effects when spin-flip is involved. In
general, interface roughness and disorder can be important quantitatively, but have been
shown not to qualitatively change the features predicted by a ballistic model.
\cite{Itoh:1999, Mathon:1999, Itoh:2003, Xu:2006} 

The free electron Stoner model is only a poor representation of the real electronic
structure of transition metals for the tunneling problem: it fails to properly reproduce
the nearly  half-metallic nature of transition metal ferromagnets based MgO tunnel
junctions, that is caused by the symmetry of the bands at the Fermi energy,
\cite{Zhang:2004} leading to the underestimated TMR ratios by our model noted above. On
the other hand, the band structure calculations by Heiliger {\it et al}.
\cite{Heiliger:2008} show that the free electron model can perform quite well as far as
the torque is concerned. We explained this in Section \ref{sec:res} by its dependence on
only one interface polarization leading to a better performance of a model that is not
accurate in this respect. The band structures calculations in Ref.
\onlinecite{Heiliger:2008} indicate that the torque is strongly localized to a few
monolayers which is in support of our simple model. 

The issue of the wave-function symmetry should also be considered when a normal metal is
inserted. When the electrons with wave vector normal to the interface dominate, the normal
metal (Cr) is actually a potential barrier for the Fe majority spins \cite{Yuasa:2008}
rather than a potential well as assumed here.



In contrast to metallic spin valve structures, in which the out-of-plane torque is
generally less than 10\% of the in-plane counterpart, the out-of-plane contribution has
been found quite large in tunneling junctions (a 30\% contribution at high bias is
measured in Ref. \onlinecite{Sankey:2008}). Close to the zero crossing of the in-plane
torque at positive bias the out-of-pane torque should become dominant. 

An experimental ``phase diagram'' that can be compared with \Figure{dc_V} would constitute
a stringent test of our predictions.  Since the barrier height and the effective electron
mass in the barrier can not be controlled, we suggest to measure the torques
systematically for several MTJ structures with different barrier width (otherwise
identical) to test the red curve in the right panel of \Figure{dc_V}. The zero crossing of
the in-plane torque is predicted to occur at voltages that are too high for the
experiments in Refs. \onlinecite{Sankey:2008, Kubota:2008, Deac:2008}. For wider tunneling
barriers it should happen at smaller voltages. 

\section{Summary} \label{sec:sum}

To summarize, scattering theory of transport is used to calculate the spin-transfer torque for
a magnetic multilayer structure at finite bias. The experimental spin-torque data (both
in-plane and out-of-plane) can be reproduced using realistic parameters in our model. The
spin-transfer torque on a given layer may change sign for only one bias polarity. The torque
zero-crossing is caused by the combined effects of the polarization sign change at the FI
interface and the focusing effect of the barrier. The bias voltage required for torque
zero-crossing decreases as a function of the barrier width. The out-of-plane torkance at zero
bias vanishes for symmetric FIF structures, but remains finite for asymmetric structures. In
FNIF structure we find on top of the previously reported oscillating TMR \cite{Yuasa:2002} and
charge pumping voltage \cite{Xiao:CP} that the spin-transfer torque also oscillates and may
change sign with the N layer thickness.

\section*{Acknowledgement}

We acknowledge M. D. Stiles, C. Heiliger, and A. Deac for helpful discussions. This work
is supported by EC Contract IST-033749 ``DynaMax''.

\appendix 
\section{Optical theorem in tunnel junction} 

The scattering matrices in \Eq{eqn:sC} include the reflection amplitude inside the
insulator ($\hat{r}_1'$ and $\hat{r}_2$). Expressions become more transparent when the
reflection amplitudes are replaced by the transmission probabilities, however. This can be
achieved by the optical theorem tailored for tunnel junction that reflects current
conservation. Note that the equivalent statement in metallic systems is the well-known
relation $\abs{r}^2+\abs{t}^2 = 1$ that follows from the unitary of the scattering matrix.

The optical theorem for light is derived from conservation of energy, whereas in
electronic transport it is based on conservation of charge. We consider here the
non-standard situation of the interface between a metal and a tunneling barrier, for which
the unitary of the scattering matrix cannot be invoked without some care. Let us consider
a non-magnetic IXN structure, where X could be basically anything. We assume
flux-normalized plane waves $e^{ik_n x}/\sqrt{k}_n$ in N with mode index $n$, and
exponential solutions $e^{\pm\kappa_m x}/\sqrt{\kappa_m}$ in I with mode index $m$. The
electron with wave function $e^{-\kappa_p x}/\sqrt{\kappa_p}$ in I is reflected with
amplitude $r_{mp}$, and transmitted into N with amplitude $t_{np}$. Then the wave
functions in I and N are:
\begin{equation}
	\psi_{\rm I}^m = {e^{-\kappa_p x}\over\sqrt{\kappa_p}} \delta_{mp}
				 + r_{mp} {e^{\kappa_m x}\over\sqrt{\kappa_m}}, \quad
	\psi_{\rm N}^n = t_{np} {e^{ik_n x}\over\sqrt{k_n }}.
\end{equation}
The current in I is given by the imaginary part of the reflection amplitudes since
\begin{equation}
	I_{\rm I} 
	= {\hbar\over m} \sum_m \im{\psi_{\rm I}^{m*}\partial_x\psi_{\rm I}^m }
	 = {2\hbar\over m} \im{r_{pp}}, 
\end{equation}
The current in N reads 
\begin{equation}
    I_{\rm N} = {\hbar\over m} \sum_n 
			\im{\psi_{\rm N}^{n*}\partial_x\psi_{\rm N}^n} 
			= {\hbar\over m}\sum_n \abs{t_{np}}^2. 
\end{equation}
By current conservation: $I_{\rm I} = I_{\rm N}$, we have
\begin{equation}
	2\im{r_{pp}} = \sum_n\abs{t_{np}}^2 \equiv \sum_n T_{np}.
\end{equation}
This relation reduces to
\begin{equation}
	2~\im{r_{pp}} = \abs{t_{pp}}^2 \equiv T_{pp}.
\end{equation}
for the ballistic model used in the text.

\section{Polarization sign change} 

\begin{figure}[b]
	\psfrag{U}{$U_{\rm b}$}
	\psfrag{Et}{$E_\perp$}
	\psfrag{Elu}{$E_\|$}
	\psfrag{T1}{$T_1$}
	\psfrag{T2}{$T_2$}
	\psfrag{Vu}{$V_\up$}
	\psfrag{Vd}{$V_\dn$}
	\includegraphics[width=6.cm]{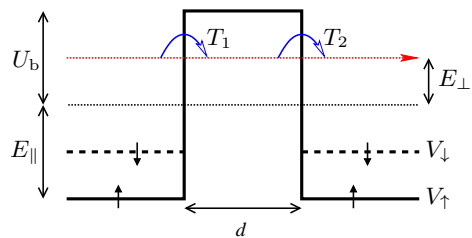}	
	\caption{(Color online) Potential barrier.}
	\label{barrier}
\end{figure}

\begin{figure}[t]
	\includegraphics[width=8.cm]{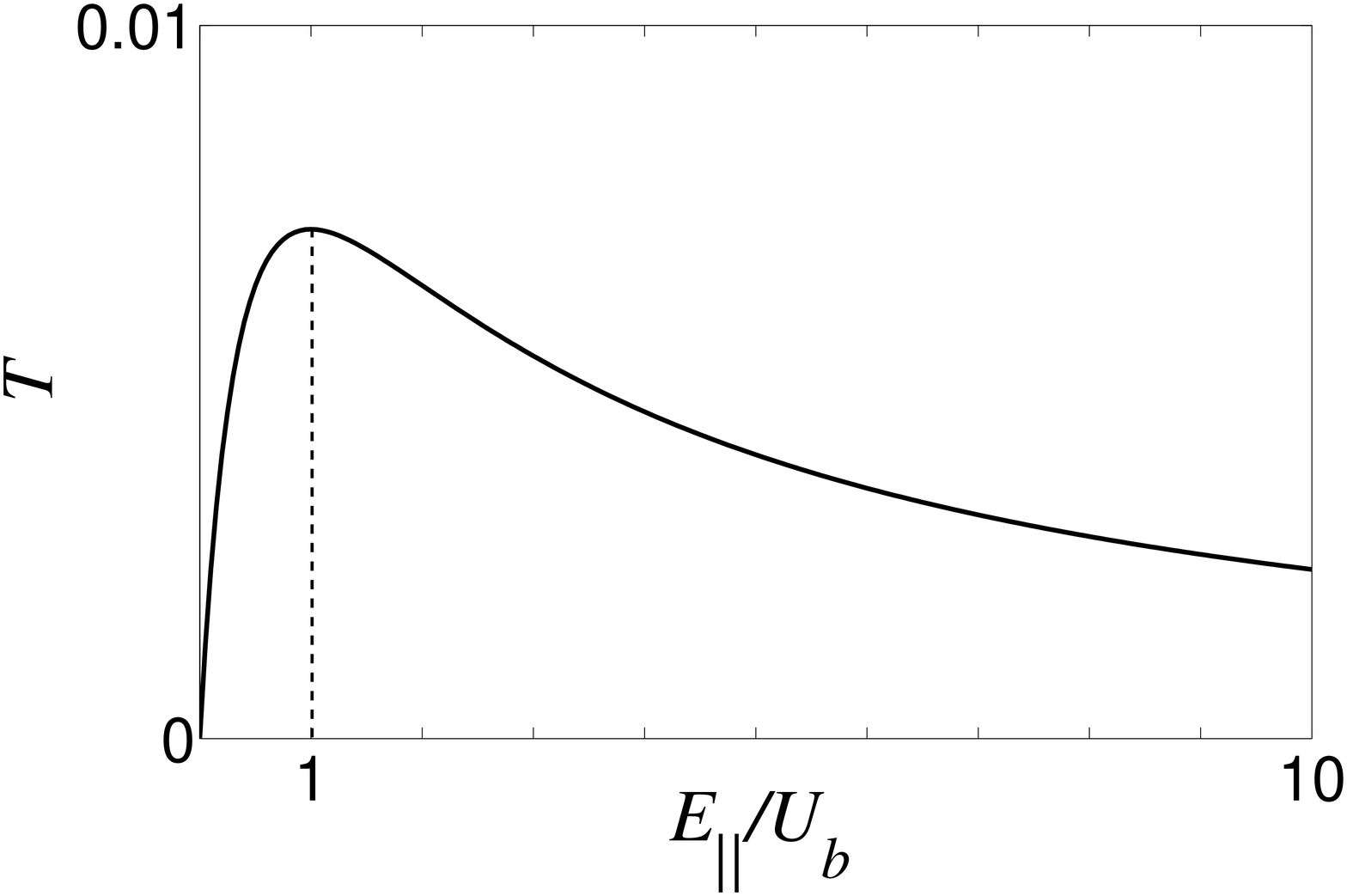}	
	\caption{$T$ vs. $E_\|/U_b$ ($U_{\rm b} = 0.25\EF, E=\EF, \beta = 0.4$, and $d = 1$ nm).}
	\label{T}
\end{figure}

For a better understanding of the polarization sign change, let us inspect the simple potential
barrier depicted in \Figure{barrier} (thick solid line), ignoring the spin dependence for the
moment. The barrier width is $d$ and the relative barrier height $U_{\rm b}
=\hbar^2\kappa^2/2m$. As seen in \Figure{barrier}, $U_{\rm b}$ is measured relative to the
longitudinal electron energy in the barrier $E-E_\perp = E_\|+V$, where $E$ is the total
electron energy and $E_\|$ and $E_\perp$ are the longitudinal (normal to the interfaces) and
transverse kinetic energy, and $V$ is the band edge. By solving this standard quantum
mechanical exercise, we find the transmission probability through the barrier
\begin{equation}
	T = {1\over 1+ {(E_\|+U_{\rm b})^2\over 4E_\| U_{\rm b}}\sinh^2(\kappa d)}
	\approx 
	{16 E_\| U_{\rm b}\over (E_\|+U_{\rm b})^2}e^{-2\kappa d},
	\label{eqn:T}
\end{equation}
where the approximation is accurate when $\kappa d\gg 1$. \Eq{eqn:T} shows that for a fixed
$U_{\rm b}$ (or $\kappa$), $T$ is maximal when $E_\|/ U_{\rm b} = 1$ (see \Figure{T}), where
$E_\|$ can be tuned by changing the band edge $V$.

In an FIF MTJ, for electrons in F with the same total energy $E =
E_\|^{\up/\dn}+V_{\up/\dn}+E_\perp$ and the same transverse energy $E_\perp$, the relative
barrier height ($U_{\rm b}$ in \Eq{eqn:T}) is the same for spin up and spin down electrons.
However, the band edges ($V_{\up/\dn}$) are spin-dependent as indicated by the solid (spin up)
and dashed (spin down) lines in \Figure{barrier}. The longitudinal kinetic energies ($E_\|$ in
\Eq{eqn:T}) is larger for spin-up than spin-down electrons. In our Stoner model we typically
find $E_\|^\dn/U_{\rm b}\sim 1$ whereas $E_\|^\up/U_{\rm b}>2$. According to \Eq{eqn:T} and
\Figure{T}, we find for the parameters in our Stoner model the surprising result that $T_\up <
T_\dn$. In general, electrons close to the Fermi energy with small transverse wave vectors $q$
show this inverted polarization sign change behaviour. For large $q$, spin-up electrons tend to
have higher transmission again, and the polarization becomes positive.


\end{document}